\documentclass{article}
\usepackage[utf8]{inputenc}

\usepackage{graphics}
\usepackage[pdftex]{graphicx,color}  
\usepackage{fancybox}
\usepackage{amsmath,latexsym,psfrag,epsf,epsfig,amssymb,rotating}
\usepackage{multirow}
\usepackage{url}
\usepackage{natbib}
\usepackage{fullpage}

\title{Bergm: Bayesian exponential random graph models in R}

\author{Alberto Caimo$^1$ and Nial Friel$^2$\\
{\small $^1$Dublin Institute of Technology, Ireland; $^2$University College Dublin, Ireland}}

\begin{document}

\maketitle

\section{Introduction}

Networks are relational data that can be defined as a collection of nodes interacting with each other and connected in a pairwise fashion. 
From a statistical point of view, networks are relational data represented as mathematical graphs. A graph consists of a set of $n$ nodes and a set of $m$ edges which define some sort of relationships between pair of nodes called dyads.
The connectivity pattern of a graph can be described by an $n \times n$ adjacency matrix $y$ encoding the presence or absence of an edge between node $i$ and $j$:

$$
y_{ij} 
  =\left\{\begin{matrix}
          1, &\textrm{if }(i,j) \textrm{ are connected,}\\
          0, &\textrm{otherwise.}\\
          \end{matrix}
   \right.
$$   
  
Two nodes are adjacent or neighbours if there is an edge between them. If $y_{ij} = y_{ji}, \forall i, j$ then the adjacency matrix is symmetric and the graph is undirected, otherwise the graph is directed and it is often called a digraph. Edges connecting a node to itself (self-loops) are generally not allowed in many applications and will not be considered in this context.

\section{Exponential random graph models}

Introduced by \cite{hol:lei81} to model individual heterogeneity of nodes and reciprocity of their edges, the family of exponential random graph models (ERGMs) was generalised by \cite{fra:str86}, \cite{was:pat96} and \cite{sni:pat:rob:han06}. 
ERGMs constitute a broad class of network models (see \citep{rob:pat:kal:lus07} for an introduction) that assume that the observed network $y$ can be explained in terms of the relative prevalence of a set of network statistics $s(y)$:
\begin{equation}
p(y | \theta) = \frac{ \exp \{ \theta^t s(y)\} }
                             {z(\theta)}
\end{equation}
where the normalising constant $z(\theta)$ is intractable for non trivially-small networks. 

Due to the complexity of networks, it is necessary to reduce the information to describe essential properties of the network. Usually this is done via network statistics, a series of counts of sub-graph configurations (e.g., the number of edges, stars, triangles, functions of degree distributions, edgewise shared parters, etc.), catching the relevant information \citep{sni:pat:rob:han06}. 

\section{Bayesian inference}

In the ERGM context (see \citep{was:pat96} and \citep{rob:pat:kal:lus07}), the posterior distribution of model parameters $\theta$ given an 
observed network $y$ on $n$ nodes maybe written as: 
\begin{equation}
p(\theta|y) 
= \frac{p(y|\theta)\; p(\theta)}{p(y)}
= \frac{\exp\{\theta^t s(y)\}}{z(\theta)}\;\frac{p(\theta)}{p(y)},
\label{eqn:bergm}
\end{equation}
where $s(y)$ is a known vector of sufficient network statistics \citep{mor:han:hun08}, $p(\theta)$ is a prior distribution placed on 
$\theta$, $z(\theta)$ is the intractable likelihood normalising constant, and $p(y)$ is the model evidence. The presence of the intractable ERGM likelihood 
implies that the usual suite of standard Bayesian inferential methods, especially standard MCMC tools are not possible in this context. However
recent work has shown that the ERGM can be given the full Bayesian treatment as we now outline. 

\section{The {\tt Bergm} package for R}

The {\tt Bergm} package \citep{cai:fri14} for {\tt R} \citep{R} implements Bayesian analysis for ERGMs \citep{cai:fri11, cai:fri13, cai:mir15, thi:fri:cai:kau16, bouranis2015bayesian}. The package provides a comprehensive framework for Bayesian inference using Markov chain Monte Carlo (MCMC) algorithms. It can also supply graphical Bayesian goodness-of-fit procedures that address the issue of model adequacy.

The package is simple to use and represents an attractive way of analysing network data as it offers the advantage of a complete probabilistic treatment of uncertainty. {\tt Bergm} is based on the {\tt ergm} package \citep{hun:han:but:goo:mor08} which is part of the {\tt statnet} suite of packages \citep{han:hun:but:goo:mor07} and therefore it makes use of the same model set-up and network simulation algorithms. The {\tt ergm} and {\tt Bergm} packages complement each other in the sense that {\tt ergm} implements maximum likelihood-based inference whereas {\tt Bergm} implements Bayesian inference.
The {\tt Bergm} package has been continually improved in terms of speed performance over the last years and we feel that this package now offers the end-user a feasible option for carrying out Bayesian inference for networks with several thousands of nodes.

\section{Approximate exchange algorithm}

In order to approximate the posterior distribution $p(\theta|y)$, the {\tt Bergm} package uses the exchange algorithm described in Section 4.1 of 
\citep{cai:fri11} to sample from the following distribution:
\begin{equation*}
p(\theta',y',\theta | y) \propto p(y|\theta)p(\theta)\epsilon(\theta'|\theta) p(y'|\theta') 
\end{equation*}
where $p(y'|\theta')$ is the likelihood on which the simulated data $y'$ are defined and belongs to the same exponential family of densities as $p(y|\theta)$, $\epsilon(\theta'|\theta)$ is any 
arbitrary proposal distribution for the augmented variable $\theta'$. As we will see in the next section, this proposal 
distribution is set to be a normal centred at $\theta$.

At each MCMC iteration, the exchange algorithm consists of a Gibbs update of $\theta'$ followed by a Gibbs update of $y'$, 
which is drawn from $p(\cdot|\theta')$ via an MCMC algorithm \citep{hun:han:but:goo:mor08}. Then a deterministic exchange or swap 
from the current state $\theta$ to the proposed new parameter $\theta'$. This deterministic proposal is accepted with probability:
\begin{equation*}
\min\left( 1, \frac{q_{\theta}(y')p(\theta') \epsilon(\theta|\theta') q_{\theta'}(y)}
{q_{\theta}(y)p(\theta) \epsilon(\theta'|\theta) q_{\theta'}(y')} 
\times \frac{z(\theta)z(\theta')}{z(\theta)z(\theta')} \right),
\end{equation*}
where $q_{\theta}$ and $q_{\theta'}$ indicate the unnormalised likelihoods with parameter $\theta$ and $\theta'$, respectively.
Notice that all the normalising constants cancel above and below in the fraction above, in this way avoiding the need to calculate the
intractable normalising constant.

The approximate exchange algorithm is implemented by the {\tt bergm} function in the following way:\\[.2cm]
\noindent{\tt for} $i = 1,\dots,N$\\[.2cm]
\vspace{.2cm}
\qquad {\tt 1. generate} $\theta'$ {\tt from} $\epsilon(\cdot|\theta)$\\ 
\vspace{.2cm}
\qquad {\tt 2. simulate} $y'$ {\tt from} $p(\cdot|\theta')$\\
\vspace{.2cm}
\qquad {\tt 3. update} $\theta \rightarrow \theta'$ {\tt (log) probability:}
$$
\min\left( 0,\; \left[ \theta - \theta'\right]^t \left[s(y') - s(y)\right]
  +\log\left[
   \frac{p(\theta')}
        {p(\theta)}\right]\right)
$$
{\tt end for}\\[.2cm]

where $s(y)$ is the observed vector of network statistics and $s(y')$ is the simulated vector of network statistics. 
Step 2. above requires a draw from the ERGM likelihood and perfect sampling in principle is a possibility, however practically this is out of reach as no such sampler has yet been developed.
Therefore the pragmatic approach we take is to run a Gibbs sampler for \texttt{aux.iters} iterations targetting $p(\cdot|\theta')$. 
In order to improve mixing a parallel adaptive direction sampler (ADS) approach \citep{gil:rob:geo94,rob:gil94} is considered as the default procedure.

To illustrate the inferential procedure, we fit a 4-dimensional ERGM to the Faux Mesa High School network data \citep{resbea97} including uniform homophily between students with the same 'grade' ({\tt nodematch('Grade')}), and statistics capturing the degree distribution ({\tt gwdegree}) and transitivity effect ({\tt gwesp}):

\begin{center}
\includegraphics[scale = 0.35]{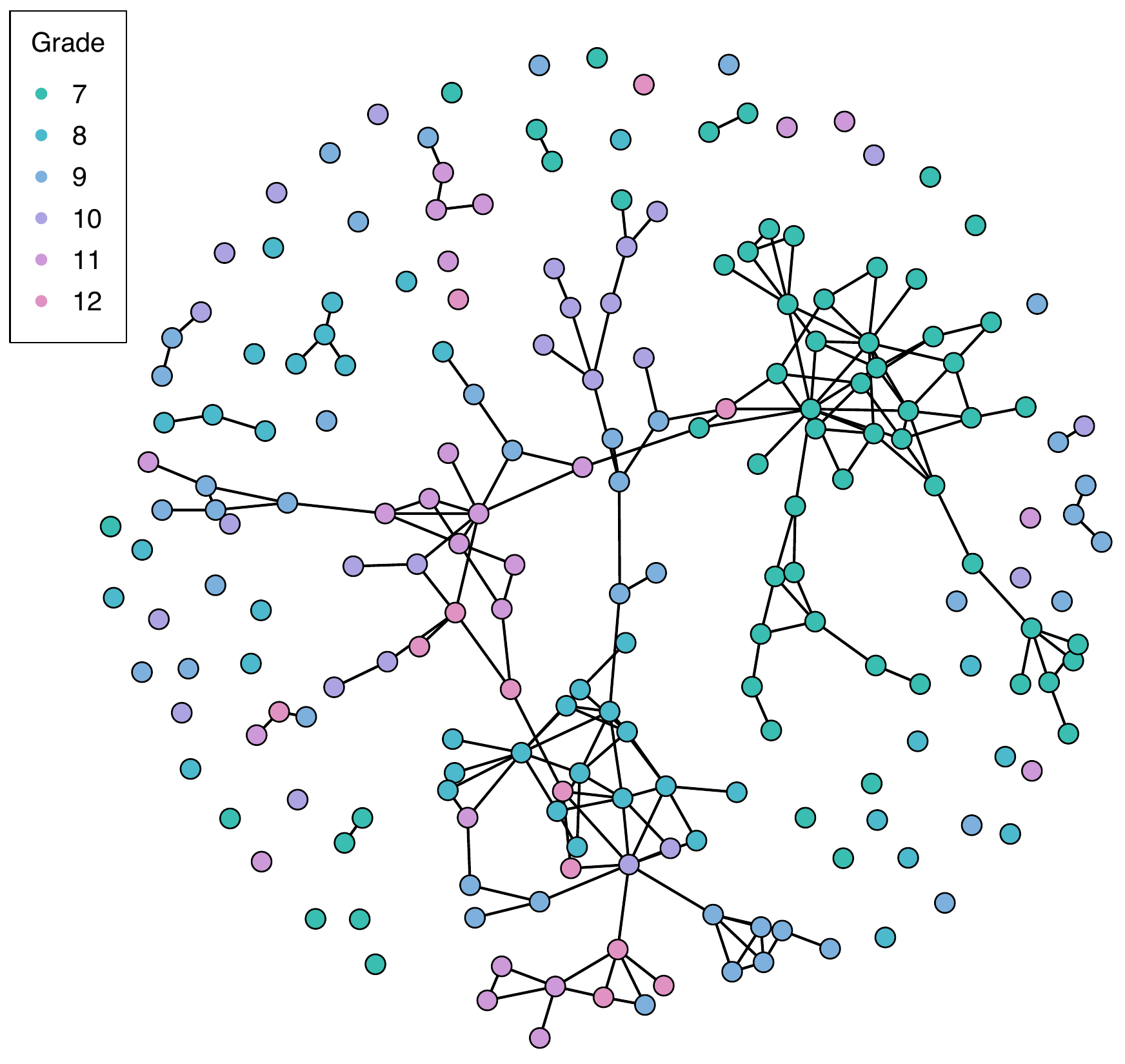}
\end{center}

\begin{verbatim}
> model <- y ~ edges + 
+              nodematch('Grade') + 
+              gwdegree(0.2, fixed = TRUE) +
+              gwesp(0.2, fixed = TRUE)  
\end{verbatim}

and we use the {\tt bergm} function with $20,000$ auxiliary iterations for network simulation and 6 MCMC chains for the ADS procedure consisting of $2,000$ main iterations each:

\begin{verbatim}
> bergm.post <- bergm(model,
+                     burn.in = 300,
+                     main.iters = 2000,
+                     aux.iters = 20000, 
+                     nchains = 6, 
+                     gamma = 0.6)
\end{verbatim}

The estimation took about 200 seconds. A summary of the MCMC results is available via the {\tt bergm.output} command:

\begin{verbatim}
> bergm.output(post)

\end{verbatim}

\begin{verbatim}
 
                               Mean        SD    Naive SE
theta1 (edges)           -6.4539945 0.2269798 0.002072032
theta2 (nodematch.Grade)  2.0653066 0.1562896 0.001426723
theta3 (gwdegree)         0.1555102 0.2156994 0.001969057
theta4 (gwesp.fixed.0.2)  1.6045295 0.1624254 0.001482734

                         Time-series SE
theta1 (edges)              0.013682987
theta2 (nodematch.Grade)    0.009303603
theta3 (gwdegree)           0.013919679
theta4 (gwesp.fixed.0.2)    0.009959682

                               2.5%        25%        50%
theta1 (edges)           -6.8988256 -6.6003492 -6.4532048
theta2 (nodematch.Grade)  1.7879473  1.9557575  2.0537140
theta3 (gwdegree)        -0.2825274  0.0158655  0.1618899
theta4 (gwesp.fixed.0.2)  1.2841624  1.4983524  1.6031113
                                75%      97.5%
theta1 (edges)           -6.3043630 -6.0255210
theta2 (nodematch.Grade)  2.1665645  2.3952768
theta3 (gwdegree)         0.3118872  0.5568012
theta4 (gwesp.fixed.0.2)  1.7099524  1.9425482

Acceptance rate: 0.1965833 
\end{verbatim}

Density and trace plots are produced automatically by the {\tt bergm.output} function:

\begin{center}
\includegraphics[scale = 0.6]{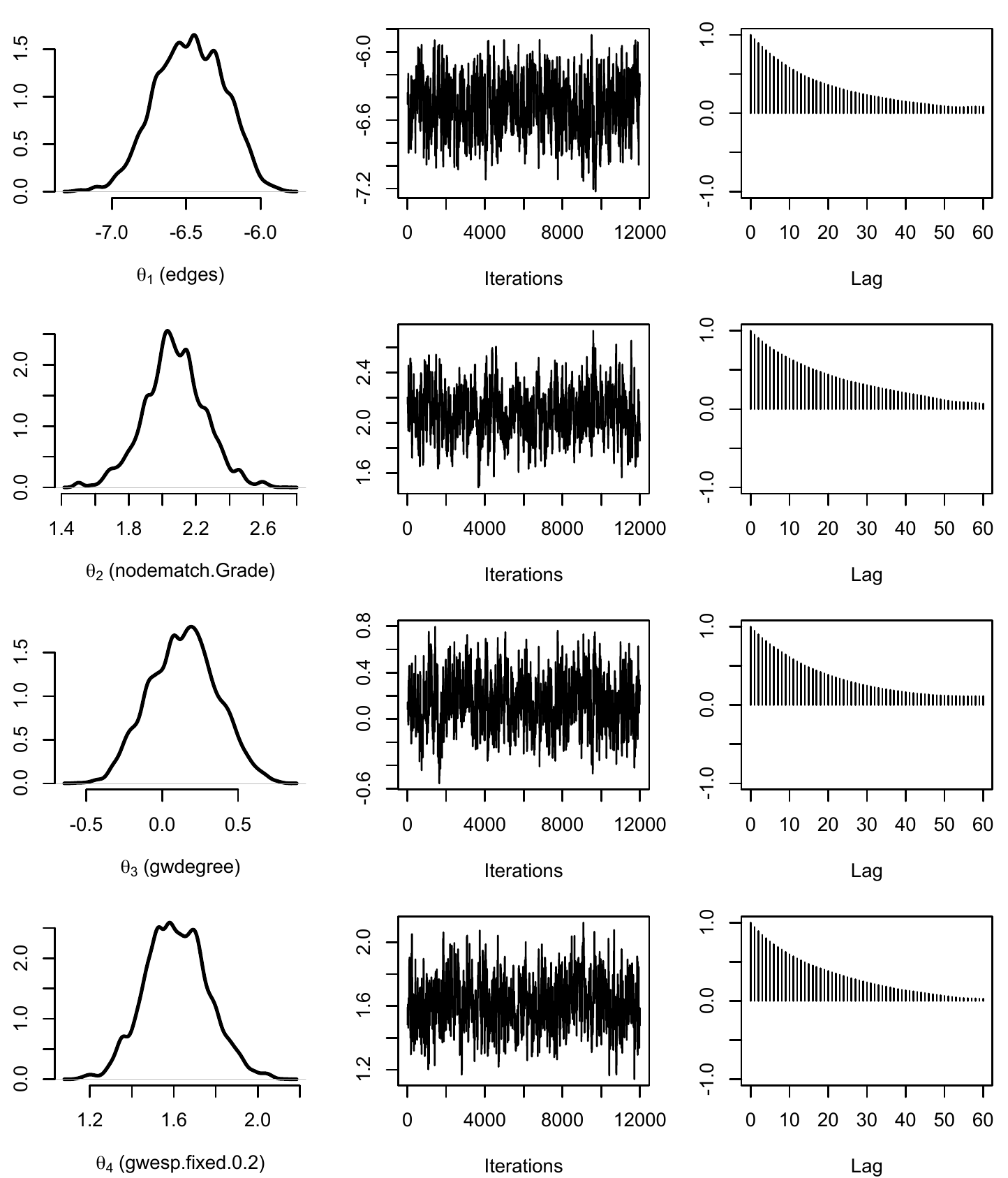}
\end{center}

Posterior predictive goodness-of-fit diagnostics plots are available via the {\tt bgof} command, as shown in the figure below. 
\begin{verbatim}
> bgof(bergm.post,
+      aux.iters = 20000,
+      n.deg = 14,
+      n.dist = 15,
+      n.esp = 10)
\end{verbatim}

\begin{center}
\includegraphics[scale = 0.7]{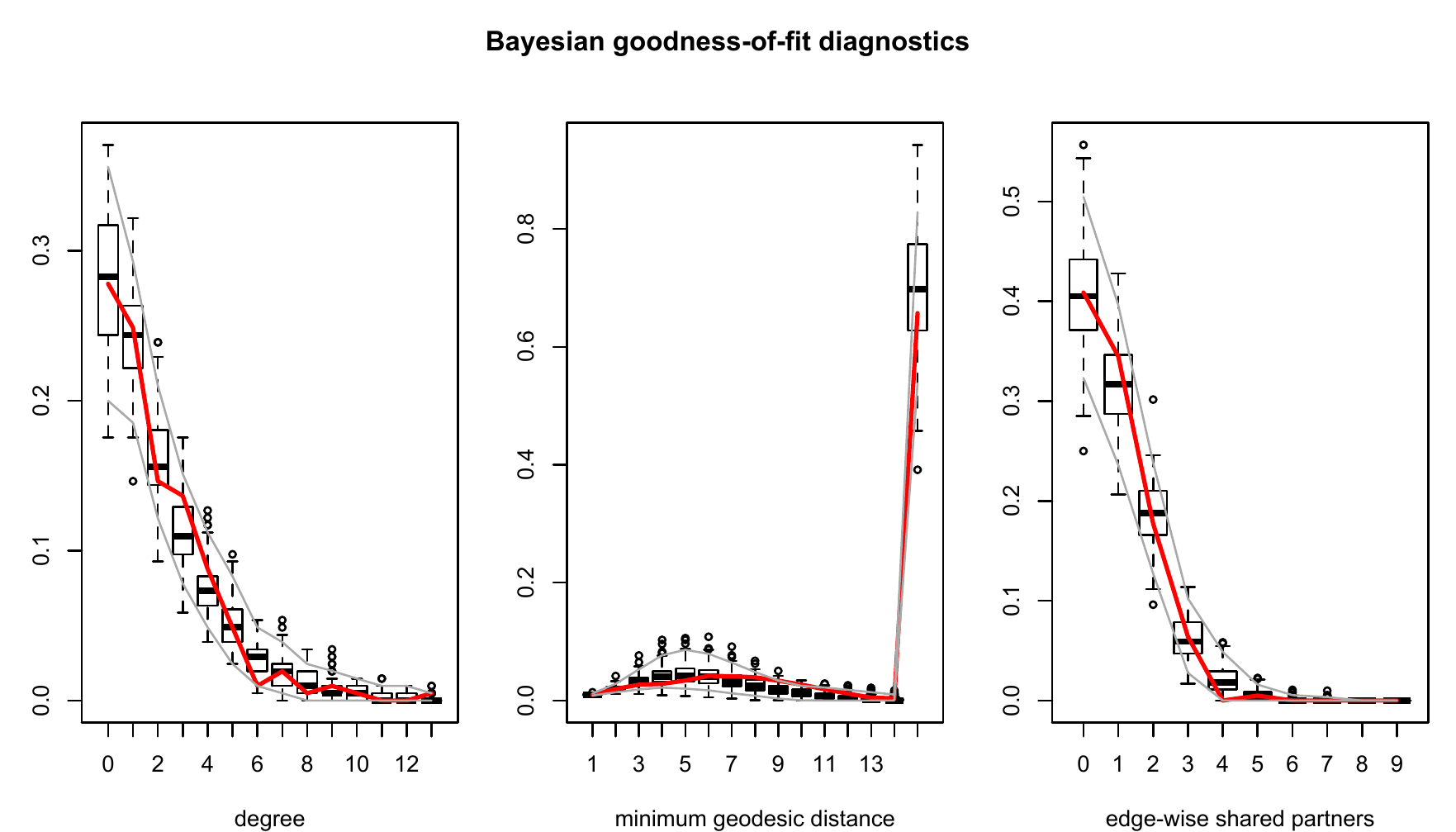}
\end{center}

The plots in the figure indicate a very good fit of the model in terms of a higher-level network statistics in the data.

\section{Pseudo-posterior calibration}

An alternative approach to Bayesian inference for ERGMs has been proposed by \cite{bouranis2015bayesian} based on replacing the intractable ERGM likelihood with a 
tractable pseudo-likelihood approximation. This results in a so-called pseudo-posterior distribution for which it is straightforward to sample from using the usual 
MCMC toolbox, for example. However it is well understood that Bayesian inference based on the pseudolikelihood can yield poor estimation and this motivated 
\citep{bouranis2015bayesian} to develop an approach which allows one to correct or calibrate a sample from 
such a pseudo-posterior distribution so that it is approximately distributed from the target posterior distribution. This is achieved by estimating the 
maximum a posteriori (MAP) of the posterior distribution and also estimating the Hessian of the posterior distribution at the MAP. Both of these quantites
can then be used to define an affine transformation of the pseudo-posterior distribution to one that is approximately distributed as the posterior distribution.

The pseudo-posterior calibration approach can be carried out using the {\tt calibrate.bergm} function:
\begin{verbatim}
> cbergm.post <- calibrate.bergm(model,
+                 iters = 1000,
+                 aux.iters = 20000,
+                 noisy.nsim = 100,  
+                 noisy.thin = 1000,
+                 mcmc = 10000)
\end{verbatim} 

The estimation took about 80 seconds and the MCMC output can be analysed by using the {\tt bergm.output} function.

In the plots below, we see that the posterior estimates from the {\tt calibrate.bergm} function is in good agreement with that corresponding to the {\tt bergm} function:

\begin{center}
\includegraphics[scale = 0.6]{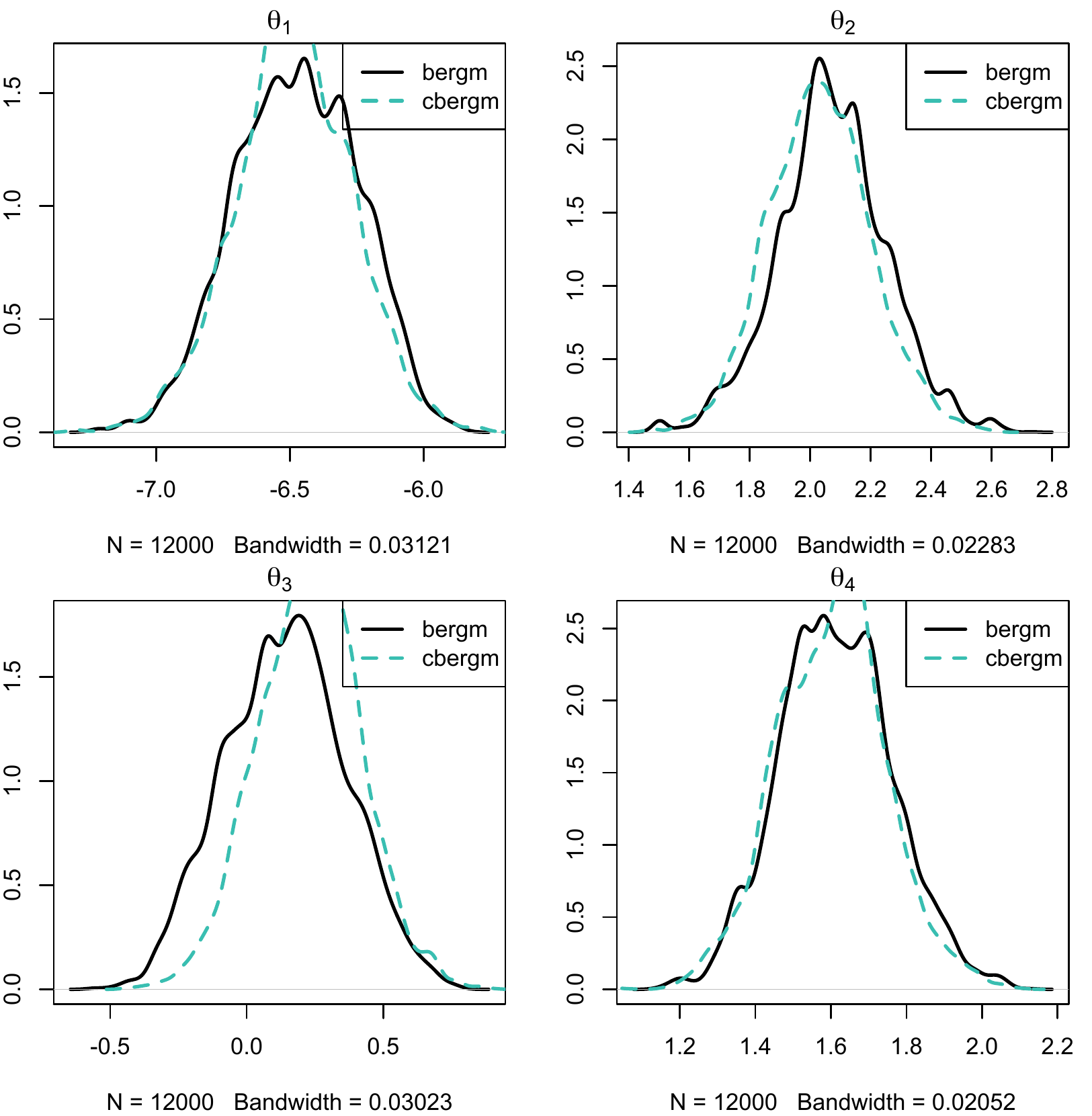}
\end{center}

\section{More information}

The {\tt Bergm} package is available on the CRAN at:
\url{https://CRAN.R-project.org/package=Bergm},
and also on GitHub at: \url{https://github.com/acaimo/Bergm}.

\bibliographystyle{asa}

\begin{thebibliography}{18}
\newcommand{\enquote}[1]{``#1''}
\expandafter\ifx\csname natexlab\endcsname\relax\def\natexlab#1{#1}\fi

\bibitem[{Bouranis et~al.(2015)Bouranis, Friel, and
  Maire}]{bouranis2015bayesian}
Bouranis, L., Friel, N., and Maire, F. (2015), \enquote{Bayesian inference for
  misspecified exponential random graph models,} \textit{arXiv preprint
  arXiv:1510.00934}.

\bibitem[{Caimo and Friel(2011)}]{cai:fri11}
Caimo, A. and Friel, N. (2011), \enquote{{B}ayesian Inference for Exponential
  Random Graph Models,} \textit{Social Networks}, 33, 41 -- 55.

\bibitem[{Caimo and Friel(2013)}]{cai:fri13}
Caimo, A. and Friel, N. (2013), \enquote{{B}ayesian model selection for exponential random graph
  models,} \textit{Social Networks}, 35, 11 -- 24.

\bibitem[{Caimo and Friel(2014)}]{cai:fri14}
Caimo, A. and Friel, N. (2014), \enquote{Bergm: {B}ayesian Exponential Random Graphs in {R},}
  \textit{Journal of Statistical Software}, 61, 1--25.

\bibitem[{Caimo and Mira(2015)}]{cai:mir15}
Caimo, A. and Mira, A. (2015), \enquote{Efficient computational strategies for
  doubly intractable problems with applications to Bayesian social networks,}
  \textit{Statistics and Computing}, 25, 113--125.

\bibitem[{Frank and Strauss(1986)}]{fra:str86}
Frank, O. and Strauss, D. (1986), \enquote{{M}arkov Graphs,} \textit{Journal of
  the American Statistical Association}, 81, 832--842.

\bibitem[{Gilks et~al.(1994)Gilks, Roberts, and George}]{gil:rob:geo94}
Gilks, W.~R., Roberts, G.~O., and George, E.~I. (1994), \enquote{Adaptive
  Direction Sampling,} \textit{Statistician}, 43, 179--189.

\bibitem[{Handcock et~al.(2007)Handcock, Hunter, Butts, Goodreau, and
  Morris}]{han:hun:but:goo:mor07}
Handcock, M.~S., Hunter, D.~R., Butts, C.~T., Goodreau, S.~M., and Morris, M.
  (2007), \enquote{statnet: Software Tools for the Representation,
  Visualization, Analysis and Simulation of Network Data,} \textit{Journal of
  Statistical Software}, 24, 1--11.

\bibitem[{Holland and Leinhardt(1981)}]{hol:lei81}
Holland, P.~W. and Leinhardt, S. (1981), \enquote{An exponential family of
  probability distributions for directed graphs (with discussion),}
  \textit{Journal of the American Statistical Association}, 76, 33--65.

\bibitem[{Hunter et~al.(2008)Hunter, Handcock, Butts, Goodreau, and
  Morris}]{hun:han:but:goo:mor08}
Hunter, D.~R., Handcock, M.~S., Butts, C.~T., Goodreau, S.~M., and Morris, M.
  (2008), \enquote{ergm: A Package to Fit, Simulate and Diagnose
  Exponential-Family Models for Networks,} \textit{Journal of Statistical
  Software}, 24, 1--29.

\bibitem[{Morris et~al.(2008)Morris, Handcock, and Hunter}]{mor:han:hun08}
Morris, M., Handcock, M.~S., and Hunter, D.~R. (2008), \enquote{Specification
  of exponential-family random graph models: terms and computational aspects,}
  \textit{Journal of statistical software}, 24, 15--48.

\bibitem[{{R Development Core Team}(2011)}]{R}
{R Development Core Team} (2011), \textit{R: A Language and Environment for
  Statistical Computing}, R Foundation for Statistical Computing, Vienna,
  Austria.

\bibitem[{Resnick et~al.(1997)Resnick, Bearman, Blum, Bauman, Harris, Jones,
  Tabor, Beuhring, Sieving, and Shew}]{resbea97}
Resnick, M.~D., Bearman, P.~S., Blum, R.~W., Bauman, K.~E., Harris, K.~M.,
  Jones, J., Tabor, J., Beuhring, T., Sieving, R.~E., and Shew, M. (1997),
  \enquote{Protecting adolescents from harm: findings from the National
  Longitudinal Study on Adolescent Health,} \textit{Jama}, 278, 823--832.

\bibitem[{Roberts and Gilks(1994)}]{rob:gil94}
Roberts, G.~O. and Gilks, W.~R. (1994), \enquote{Convergence of Adaptive
  Direction Sampling,} \textit{Journal of Multivariate Analysis}, 49, 287--298.

\bibitem[{Robins et~al.(2007)Robins, Pattison, Kalish, and
  Lusher}]{rob:pat:kal:lus07}
Robins, G., Pattison, P., Kalish, Y., and Lusher, D. (2007), \enquote{An
  Introduction to Exponential Random Graph Models for Social Networks,}
  \textit{Social Networks}, 29, 169--348.

\bibitem[{Snijders et~al.(2006)Snijders, Pattison, Robins, and
  S.}]{sni:pat:rob:han06}
Snijders, T. A.~B., Pattison, P.~E., Robins, G.~L., and S., H.~M. (2006),
  \enquote{New Specifications for Exponential Random Graph Models,}
  \textit{Sociological Methodology}, 36, 99--153.

\bibitem[{Thiemichen et~al.(2016)Thiemichen, Friel, Caimo, and
  Kauermann}]{thi:fri:cai:kau16}
Thiemichen, S., Friel, N., Caimo, A., and Kauermann, G. (2016),
  \enquote{Bayesian exponential random graph models with nodal random effects,}
  \textit{Social Networks}, 46, 11--28.

\bibitem[{Wasserman and Pattison(1996)}]{was:pat96}
Wasserman, S. and Pattison, P. (1996), \enquote{Logit Models and Logistic
  Regression for Social Networks: I. {A}n Introduction to {M}arkov graphs and
  $p^*$,} \textit{Psycometrica}, 61, 401--425.

\end{thebibliography}

\end{document}